# Research on the fast Fourier transform of image based on GPU

Feifei Shen     Zhenjian Song     Congrui Wu

Jiaqi Geng      Qingyun Wang

**Abstract:** Study of general purpose computation by GPU (Graphics Processing Unit) can improve the image processing capability of micro-computer system. This paper studies the parallelism of the different stages of decimation in time radix 2 FFT algorithm, designs the butterfly and scramble kernels and implements 2D FFT on GPU. The experiment result demonstrates the validity and advantage over general CPU, especially in the condition of large input size. The approach can also be generalized to other transforms alike.

Fourier transform is a common method for image analysis，widely used in image filtering, image recognition and other fields. When the image size is large, Fourier transform is of great computation., A 1024 x 1024 image using the fast Fourier transform (FFT) requires $2.1 * 10^7$ floating-point operations. The micro computer system generally adopts the way of CPU plus DSP to solve the problem of real-time image processing. The graphics processor (GPU) of the graphics card has a strong floating-point computing power, uses it to image, graphics processing do not need to develop complex, expensive hardware can get higher performance, So GPU calculation is a promising way of calculation.

## 1.GPU programming mode

Early graphics processing pipeline is the curing is not programmable, so people can't use GPU for general calculation. In order to enhance graphics rendering capabilities, Nvidia and ATI and other companies in 2003 introduced a programmable GPU graphics card, have gained great success and keep the momentum of high speed development, such as Nvidia Geforce8800 2006 series of products chip integrated 128 programmable stream processor, floating point operation ability of up to 500GigaFlops, far higher than the current microcomputer general CPU. Many scholars have studied the graphics processor for general purpose computing, which is called GPGPU. Different from general-purpose processors: graphics processors are designed for rendering, so it should be considered in common when considering its special features.

### 1.1 Input and output

For general-purpose processor, input and output data are stored in a computer's main memory, the processor can directly access the data. But the graphics processor can access the memory card. The input data GPU general computing must be packaged into some sort of graphical data through an application sent to the memory interface from the main memory, the end of the operation, and the results also in the form of graphic data from memory read back to the main memory. Experimental results show that: the data transfer rate between the main memory and memory constraints GPU general computing is currently a major bottleneck.

## 1.2 Operation mode

Another important difference between the GPU and GPU general-purpose processor is a "stream computing" in parallel mode, all executed concurrently GPU shader function that is executed when rendering (Shader Program), graphical data flows to these parallel GPU, through colored rendering results to calculate the output function, the coloring function is also known as GPU computing cores. In order to adapt to "stream computing" model, The general computation algorithm should be programmed to compute the kernel of the graphics processor, and then execute the appropriate drawing commands, the calculation result is rendered graphics.

## 2. FFT theory analysis

Since GPU works in parallel, it is necessary to analyze the parallelism of FFT.. A N-point (N for a power of 2) sequence x (n) is defined for DFT:

$$X(k) = \sum_{n=0}^{N-1} x(n) W_N^{nk} \quad (1)$$

The computational complexity of the computation is O (N2), and the X (n) is divided into the sequence Xe and the odd sequence XO, then there are

$$X(k) = \sum_{r=0}^{N/2} x_e W_{N/2}^{rk} + \sum_{r=0}^{N/2} x_o W_{N/2}^{rk} = X_{ie}(k) + W_N^k X_{io}(k), k = 0,1...\frac{N}{2}-1 \quad (2)$$

Xie (k) and Xio (k) are the Fourier transform for XO and Xe, respectively.. (2) the X (k) of the first N/2 term is calculated, and the periodicity of the rotation factor can be obtained after the calculation of the N/2 term.:

$$X(k+\frac{N}{2}) = X_{ie}(k) - W_N^k X_{io}(k), k = 0,1...\frac{N}{2}-1 \quad (3)$$

So a N-point DFT transform into two N / 2 point DFT, and the amount of calculation is about the half of the original. According to this method, Xe and XO can be decomposed the log (n) times until the last n = 2 dft. After such decomposition, Fourier transform computation complexity decreased to O(n * log (n)), this is the basic principle of base-2-decimation in time-domain FFT It can be described with famous butterfly signal flow graph as:

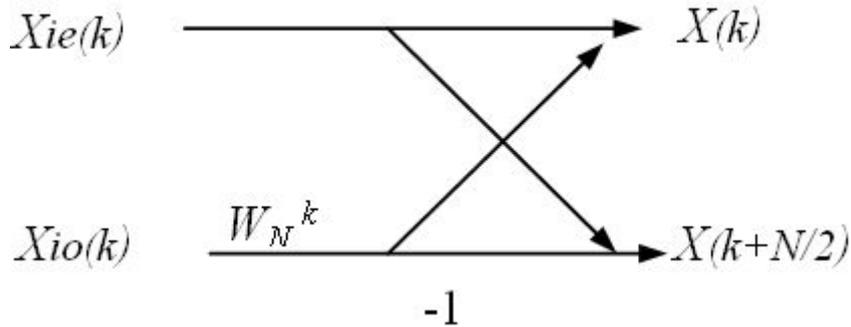

Figure 1 FFT butterfly flow chart

The computation of FFT point requires the log(N) - level butterfly operation, and it needs a N/2 butterfly operation at each level.Multidimensional Fourier transform is separable, a multidimensional Fourier transformation can be decomposed into multiple one-dimensional Fourier transform, each of the one-dimensional Fourier transform are for a variable, such as a two-dimensional Fourier transform can be achieved by two one-dimensional transform.

From the principle of fast Fourier transform, we can see that in Fourier transform level and levels of the butterfly operation are serial, and it is necessary to find out the current node for the next level of butterfly operation; but each level within a single butterfly operation is independent, which can be parallel computed .In the two-dimensional Fourier transform, the sequence of the row and column transform can be row after column or row before column, the row and column transform is serial, When the transform in each row or each column is completely independent .

## 3 .GPU implementation of two-dimensional FFT

Many scholars and developers have tried to use GPU to achieve fast Fourier transform, the more successful is the GPUFFTW project of Carolina University in North America, which have realized the one-dimensional Fourier transform based on GPU.

It uses the Autosort method of Stockham, of which the operation speed is fast and the maximum is up to 29GigaFlops. But GPUFFTW can only make one transformation, but can not make the multidimensional transformation; in addition, the software does not fully support for all the cards in the market.

Another successful library is CuFFT of Nvidia, it is a dedicated GPGPU library of Nvidia, which can make two-dimensional FFT calculation, but only running on its company's graphics card. The design of this paper is based on OpenGL 3D graphics library, the data transmission between main memory and graphics memory is using OpenGL texture. FFT calculation can be abstracted as two parallel computing: Butterfly calculation and reverse calculation.

### 3.1 Core design of butterfly computation

The FFT parallel analysis shows that the single stage butterfly operation of a one-dimensional FFT has the finest "parallel granularity", we can analyze its computational formula, and design its core operation.The butterfly operation expressed by (2) (3) formula can be expressed as following formula:

$$X(k) = X_1(k) + W \cdot X_2(k) \qquad (4)$$

At the start of the butterfly operation is the original image data and in the subsequent operations they store each stage butterfly operation results; W is a rotation factor which is related to the complex coefficients. if the input points n is known, it can be calculated in advance, and in the operation of the whole butterfly shaped process it will not be changed. The butterfly operation core designed accroding to the (4)is shown in Figure 2, it is obtained from the input texture data, and after a plural multiplication and a plural addition, the results would be written in the output texture. The design has two input textures:image data texture and secondary texture, types are rectangular textures of 4 components (RGBA), for a M × N image, the image texture size data is M × N, storing intermediate process data of image and the butterfly operation in R, G the two components, secondary texture is log (N) rows and N columns, each row stores the real and imaginary parts of primary rotating factor. Thus, each time we render, we need to achieve a butterfly operation, the FFT butterfly calculation of the whole image requires log (M) + log (N) times rendering.

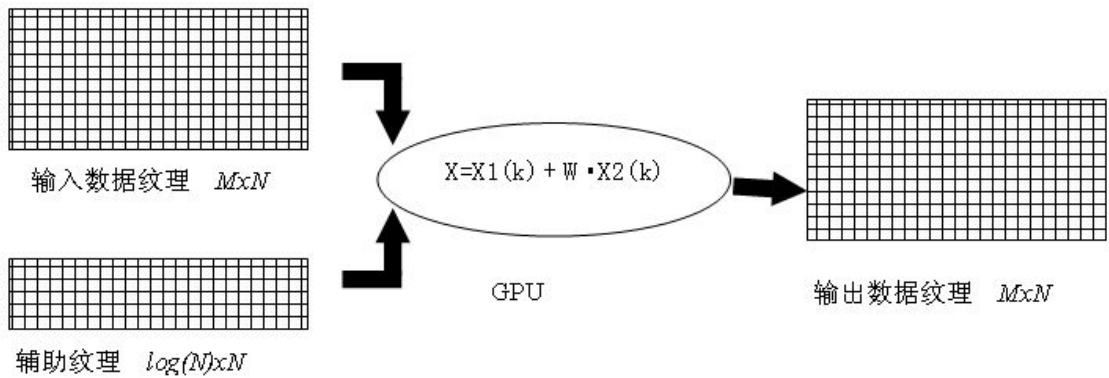

Figure 2　Butterfly operation core

## 3.2 Descending module

Site operation can reduce the storage space required for FFT, in order to achieve the site operator, Reverse operation should be made before the butterfly operation, that is inputting sequence as binary number's anti-bit sequence rearrangement. Standard descending algorithm is a single cycle, not suitable for complete using GPU. But between the two one-dimensional transform, the two-dimensional FFT must be made in reverse order, so GPU achieve descending can ensure that the entire process of FFT computation is running in the GPU and memory range, reduce the number of read data from memory back to main memory, which saves computation time. Similar to the butterfly operation's rotation factor, N is a known case, descending result indexes can be calculated before and stored in a secondary texture, descending index is obtained from secondary texture when descending rendering, and then rearranges image data based on these indexes.

## 4. Experiment Result

This experiment use Nvidia' Geforce 8600GT video card as the hardware of GPU, which has 32 stream processors ,the frequency is 540M and the driver version is 169.21.CPU is P4 dual core whih belongs to INTEL's 3G frequency. Software development tool is VC6.0, using Intel OpenCV library as the CPU test environment, after optimization the library performance is very excellent.Using this paper and OpenCV library for different sizes of gray images for fast fourier transform, the obtained results are as table 1.

| Image Size | OpenCv Time | GPU Time | GPU Maximum Error |
|---|---|---|---|
| 128×128 | <1ms | 16ms | 0.007 |
| 256×256 | 16ms | 16ms | 0.007 |
| 512×512 | 48ms | 28ms | 0.007 |
| 1024×1024 | 200ms | 62ms | 0.007 |
| 2048×2048 | 1000ms | 250ms | 0.007 |

Table 1 Comparison of CPU and GPU

It can be say,when the image size is small, using CPU directly for FFT is faster than using GPU. when the image size is less than 128 x 128, using CPUfor FFT is less than 1ms, while the GPU need about 16ms.However when the image size becomes larger, the advantage of GPU is obvious. In order to test the performance of this paper, compared it with GPUFFTW and CuFFT library, it can be seen that the performance is satisfactory by using GPU for FFT with different size images.

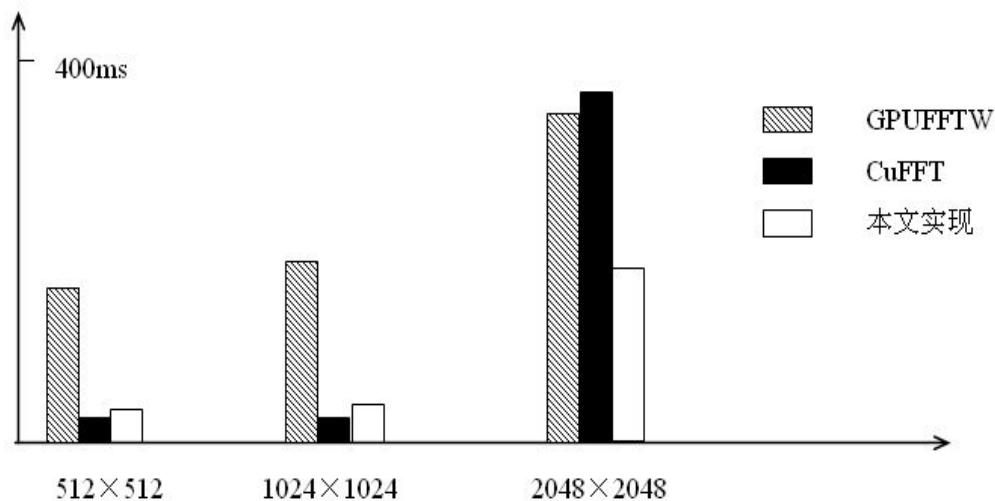

Figure 3 Comparison of several GPU implementations

Although GPU has a great potential in realizing FFT, there are also some deficiencies.:

(1) At present,GPU can only support single - precision floating - point operations, so the accuracy of the computation is not necessarily satisfied with the application requirements.

(2) AS the texture size has an upper limit, so image size for FFT is also limited, it often related with the number of the video cards memory.

## 5 .Summary

In this paper, we put up a new method of realizing image FFT based on GPU ,which has advantages of simple hardware structure and high speed, and provides a new way for real-time image processing of microcomputer system..Using the method and the idea of this paper, it can be easily implemented fourier verse transform and discrete cosine transform and other similar mathematical transformation.

The innovations of this paper are as follows:we compliy FFT kernel by using OpenGL Shading Language,The format of the corresponding texture data and auxiliary texture format are also designed,the two-dimensional FFT on the graphics processor is achieved,too.This method is fast and can be easily ported to a variety of operating system platforms with little texture, suitable for any video card that supports

OpenGL standard 2.0 and has a good portability,it can be easily transplanted to a variety of operating system platforms.